\documentclass{aa}
\usepackage{psfig,latexsym,longtable,lscape}
\def\ros{{\sl ROSAT}}

\newcommand{\D}{$^\circ$}
\def\p0{\phantom{0}}
\newcommand\approxlt{\mbox{$^{<}\hspace{-0.24cm}_{\sim}$}}
\newcommand\approxgt{\mbox{$^{>}\hspace{-0.24cm}_{\sim}$}}

\def\it{\sl}

\begin{document}
           
   \thesaurus{11         
              (04.03.1;  
               11.09.1;  
               11.01.2;  
               11.09.4;  
               13.18.1;  
               13.25.2)  
             }
   \title{Probing the gas content of the dwarf galaxy NGC~3109 with background X-ray sources}
 
   \author{P. Kahabka\inst{1},
           T. H. Puzia\inst{1},
           W. Pietsch\inst{2}
          }

   \offprints{pkahabka@astro.uni-bonn.de}
 
   \institute{$^1$~Sternwarte, Universit\"at Bonn, 
              Auf dem H\"ugel 71, 53121 Bonn, Germany\\
              $^2$~Max-Planck-Institut f\"ur extraterrestrische Physik,
              D--85740 Garching bei M\"unchen, Germany
             }

   \date{Received 10 April 2000 / Accepted 23 May 2000}
 
   \maketitle\markboth{P. Kahabka et al.: Probing the gas content of NGC~3109}
                      {P. Kahabka et al.: Probing the gas content of NGC~3109}

   \begin{abstract}
     We established the catalog of X-ray point sources in the field of 
     the Magellanic-type spiral galaxy NGC~3109 (DDO~236) from two 
     {\sl ROSAT PSPC} observations. Of the 91 X-ray sources 26 are 
     contained within the H{\sc i} extent of NGC~3109 as derived by Jobin \& 
     Carignan (1990) with the VLA. For 10 of these we can determine accurate 
     hardness ratios $\delta H\!R2\le$0.2. We find 3 candidate AGN, 2
     candidate X-ray binaries and one source which may belong to either
     class. We also find 2 candidate foreground stars. In a field of
     8\arcmin~$\times$~8\arcmin\ observed with the {\sc NTT} in the I-band 
     and centered on the nucleus of NGC~3109 we determine candidates for
     optical counterparts in the X-ray error circle of 7 {\sl ROSAT PSPC} 
     sources. We apply a spectral fit to the {\sl ROSAT} spectrum of the 
     X-ray brightest absorbed candidate AGN behind NGC~3109, RX~J1003.2-2607. 
     Assuming a galactic foreground hydrogen column density of 
     $\rm 4.3\ 10^{20}\ cm^{-2}$ we derive from the X-ray spectral fit, 
     assuming reduced metallicities ($\sim$0.2 solar), a hydrogen column 
     density due to NGC~3109 of $\rm 11\pm^{7}_{5}\ 10^{20}\ cm^{-2}$. This 
     value is slightly larger than the hydrogen column density derived from 
     the 21-cm observations of $\rm \sim8\ 10^{20}\ cm^{-2}$. We estimate 
     that the molecular mass fraction of the gas is not larger than 
     $\sim60$\%. 
%
%
   \keywords{Catalogs -- galaxies: individual: NGC~3109 (DDO~236) -- 
             galaxies: active -- galaxies: ISM --
             radio continuum: galaxies -- X-rays: galaxies}
   \end{abstract}

%
%
\section{Introduction}
NGC~3109 (DDO~236), a late-type dwarf spiral galaxy, has been classified 
as Sm~IV (Sandage \& Tammann 1981) and is seen almost edge-on with an 
inclination close to 80\D\ (Carignan 1985). Since this galaxy has no 
nucleus it has been assigned the morphological type Ir (van den Bergh 
1999). It is a southern Magellanic dwarf galaxy well resolved into stars 
and is one of the largest Magellanic dwarfs close to the Local Group at 
a distance of 1.36$\pm$0.10~~Mpc (Musella et al. 1997) with an apparent 
major-axis diameter of 30\arcmin\ (12~kpc) and minor-axis diameter of 
6\arcmin\ (Demers et al. 1985). 

NGC~3109 has a similar dimension as the LMC. With an absolute magnitude 
comparable to the SMC of $M_{\rm B} = -15.7$~mag it is rather underluminous. 
Estimates of the total mass run from 0.6 to $\rm 1.6\ 10^{10}\ M_{\odot}$. 
The galaxy is surrounded by a huge H{\sc i} envelope, quite larger than 
its optical size (Materne 1980; Huchtmeier et al. 1980; Jobin \& Carignan 
1990). Jobin \& Carignan (1990) derive an H{\sc i} mass of 
$\rm 5\pm1\times10^{8}\ M_{\odot}$. The rotation curve requires a large 
dark matter halo.

On deep exposures NGC~3109 exhibits spiral structure. Associations of knots 
of stars are visible along spiral arms. A globular cluster search revealed 
ten candidates. Bright blue stars are found evenly distributed over the 
face of the galaxy but particularly along the spiral arms, indicating that 
star formation is taking place on a galaxy-wide scale (Demers et al. 1985). 
The metallicity of NGC~3109 has been found to be low, similar to the SMC
(Richer \& McCall 1995). Minniti et al. (1999) establish the existence of
an extended halo of old and metal-poor stars.

NGC~3109 belongs as most luminous member to a subgroup of the Local Group
dwarfs which is relatively isolated (Mateo 1998). Another member of this 
group is the dwarf spheroidal galaxy Antila at a distance of only 1.\D2 
from NGC~3109. From this apparent separation on the sky the lower limit 
to the distance between both galaxies is $\sim$26~kpc (Whiting et al. 1997).

In this article we derive the X-ray population in the field of NGC~3109 from
{\sl ROSAT PSPC} observations. We classify a few of the X-ray sources 
which coincide with the H{\sc i} extent of NGC~3109 from their X-ray spectral
properties. We find two candidate AGN which we use to probe the gas content
of NGC~3109.

\section{Observations}
 
The observations used were carried out with the {\sl PSPC} detector of the 
{\sl ROSAT} observatory during two pointed observations in May-June and 
November~1992. The satellite, X-ray telescope (XRT) and the focal plane 
detector ({\sl PSPC}) are described in detail in Tr\"umper (1983) and 
Pfeffermann et al. (1986). We note that no observations have been performed 
in the direction of NGC~3109 with the {\sl HRI} detector of {\sl ROSAT}.

\begin{table*}[htbp]
  \caption[]{Data of pointings on NGC~3109}
  \begin{flushleft}
  \begin{tabular}{lcccc}
  \hline
  \noalign{\smallskip}
   Sequence & RA         & Dec  & Time interval & Expos. \\
   Number   &\multicolumn{2}{c}{(J2000.0)}   &   &(ksec)  \\
  \noalign{\smallskip}
  \hline
  \noalign{\smallskip}
  600174p &10$^h$3$^m$7.$^s$2&-26\D9$\arcmin$36$\arcsec$&26-May-92 10:14 -- 10-Jun-92 17:27 & 18.5  \\
  600385p &10$^h$3$^m$7.$^s$2&-26\D9$\arcmin$36$\arcsec$&18-Nov-92 02:07 -- 18-Nov-92 21:27  & 13.4 \\
  \noalign{\smallskip}
  \hline
  \end{tabular}
  \end{flushleft}
  \label{tab:pointing}
\end{table*}

\begin{figure}[htbp]
  \centering{ 
  \vbox{\psfig{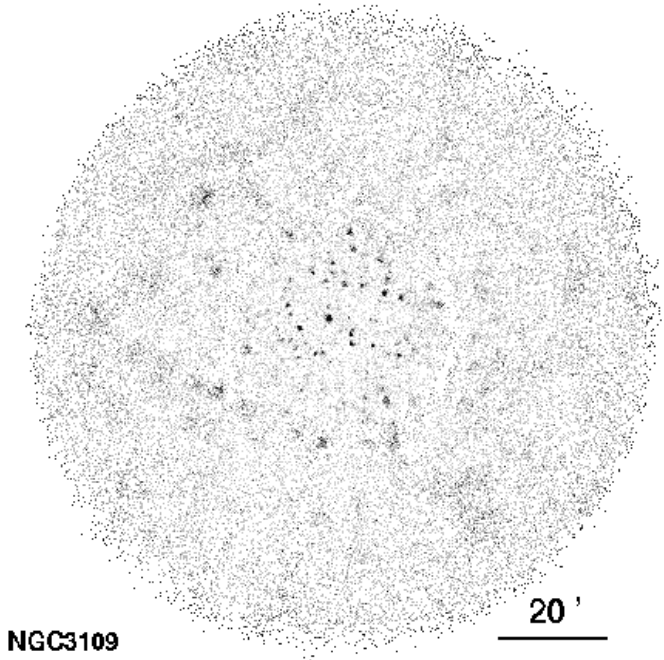}}\par
            }
  \caption[]{Exposure corrected {\sl ROSAT PSPC} image of the NGC~3109 
             field in the hard band (0.5 -- 2.0 keV). North is up and 
             east is to the left.\label{ps:figros}}
\end{figure}

The data of the two pointings given in Table~\ref{tab:pointing} have been 
retrieved from the public {\sl ROSAT} archive. The data sets from the two 
observations have been merged to one data set using standard {\sl EXSAS} 
procedures (Zimmermann et al. 1994).

\section{The catalog of X-ray sources}

Three detection procedures (local, map, and maximum likelihood) were applied 
to the merged pointings using {\sl EXSAS} commands (Zimmermann et al. 1994). 
In the local and map detection a square shaped detection cell is slid over
the image. Source counts are determined within the cell, background counts
either from an area surrounding the cell (local background) or from a 
(smoothed) background image within the same cell. In the maximum likelihood 
detection the distribution of the detected photons above the background is
compared in a maximum likelihood ratio test with the analytical point-spread
function.

The analysis was performed in the five energy channel ranges Soft = (channel 
11-41, 0.1-0.4~keV), Hard = (channel 52-201, 0.5-2.1~keV), Hard1 = (channel 
52-90, 0.5-0.9~keV) and Hard2 = (channel 91-201, 0.9-2.0~keV) and broad
(0.1-2.4~keV). The five source lists were merged to one final source list 
taking detections at off-axis angles $\le$50\arcmin\ into account. The maximum
likelihood algorithm was used to determine the final source position, the 
counts in five energy bands and the source extent. A one-dimensional energy 
and position dependent Gaussian distribution was applied in order to obtain 
the source extent. The source extent ($Ext$) is given as the Gaussian 
$\sigma_{\rm Gauss}$

\begin{equation}
  Ext = \sigma_{\rm Gauss} = FWHM_{\rm Gauss} / 2.35
\end{equation}

Hardness ratios $H\!R1$ and $H\!R2$ were calculated from the counts in the
bands as $H\!R1=(H-S)/(H+S)$ and $H\!R2=(H2-H1)/(H1+H2)$.
The existence likelihood ratio and the extent likelihood ratio was
calculated according to Cash (1979) and Cruddace et al. (1988).
We selected for our final source catalog only detections with an existence 
likelihood ratio \mbox{$LH_{\rm exist}\ge10$}, which is equal to a probability
of existence \mbox{$P\sim(1-\exp(-LH_{\rm exist}))\sim(1-4.5\times10^{-5})$}. 
We give the value for the extent only in case the extent likelihood ratio is 
\mbox {$LH_{\rm ext}\ge20$}.
A $90\%$ source error radius was calculated, adding quadratically
a systematic error of $5$\arcsec\ (cf. K\"urster 1993). 

\begin{equation}
  P_e = 2.1\times \sqrt{x_{\rm err}^2 + y_{\rm err}^2 + (5\arcsec)^2}
\end{equation}

The positional error derived for large off-axis angles \mbox {$\Delta
  \approxgt30'$} may be somewhat underestimated due to the asymmetry
of the point-spread-function. But the positional error should not be
larger than $\sim$1\arcmin. 

We finally screened the catalog of {\sl ROSAT} sources by displaying the
positions of these sources on the hard, soft, and broad band {\sl ROSAT}
{\sl PSPC} image. We could confirm 91 of the detected sources. This screened 
catalog of pointlike and moderately extended sources is given in 
Table~6.
\footnote{Table~6 will be made available on-line with the 
electronically published version.}
We give in the first column of the catalog for each confirmed source the 
sequence number and in the second column the 
source number (from the catalog of unscreened detections). We always refer
to the source number in the text.

\begin{figure*}[htbp]
  \centering{ 
  \vbox{\psfig{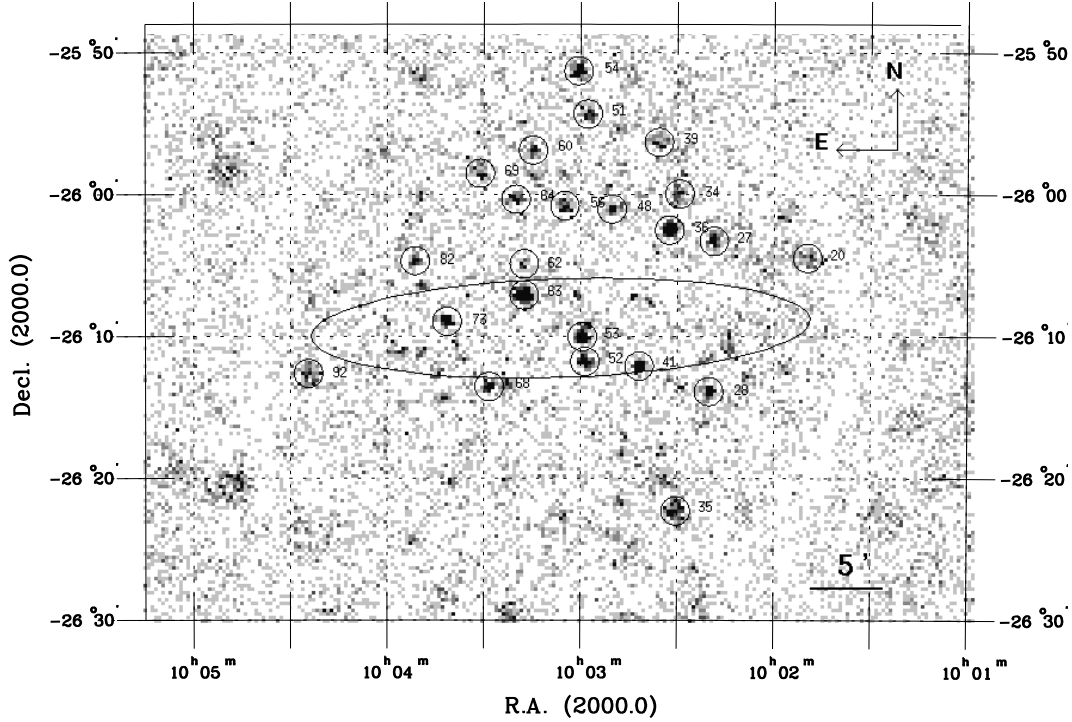}}\par
            }
  \caption[]{{\sl ROSAT PSPC} image (0.5-2.0 keV) of NGC~3109. The 22
             sources with accurate hardness ratios $H\!R2$ (and in addition 
             the AGN candidate 62) are marked with circles and the source
             number is given. Also shown is the ellipse of the stellar 
             surface density profile of this galaxy (Demers et al. 1985).
             \label{ps:rossou}}
\end{figure*}

In Fig.~\ref{ps:figros} we show the {\sl ROSAT} {\sl PSPC}\ image of the 
NGC~3109 field in the hard band (0.5 -- 2.0 keV). We note that the dwarf 
spheroidal galaxy Antila which is $\sim$1.\D2\ to the south of NGC~3109 is 
outside of the observed field of view. In Fig.~\ref{ps:rossou} we mark on 
the {\sl ROSAT PSPC} image of the central 20\arcmin\ of NGC~3109 the sources 
for with accurate hardness ratios $\delta H\!R2\le$0.2 have been determined.

\section{Classification of the X-ray sources}

The X-ray colors (hardness ratios $H\!R1$ and $H\!R2$) as given in 
Table~6 can be used for a source classification. Kahabka 
et al. (1999, hereafter KPFH99) have made a classification of the 
{\sl ROSAT PSPC} X-ray sources in the field of the Small Magellanic Cloud 
(SMC). X-ray binaries have on average harder X-ray spectra than supernova 
remnants and background AGN. The range of hardness ratios $H\!R1$ and $H\!R2$ 
is then different. In addition absorption due to intervening cold gas affects 
the X-ray colors. KPFH99 have derived the dependence of $H\!R1$ and $H\!R2$ 
on the absorbing column density for a low metallicity galaxy like the SMC. 
The hydrogen column density in the direction of the X-ray source has been 
derived from the high-resolution 21-cm map of Stanimirovic et al. (1999). 
Then from the measured X-ray color $H\!R2$ a classification could be obtained.
The strength of this method is that X-ray binaries and AGN cover different 
parts of the diagram for intervening hydrogen columns $\rm \approxlt6\times10^{21}\ cm^{-2}$. AGN are located in a band in this diagram with radio loud AGN 
populating the upper regime of the band while radio quiet AGN occupy the 
lower part of the band (cf. Laor et al. 1997; Brinkmann et al. 1997). 

For radio loud AGN spectra (powerlaw photon index $-2.0$) the dependence of 
the X-ray colors $H\!R1$ and $H\!R2$ on the hydrogen column density 
$N_{\rm H}$ ($\rm 10^{21}\ {\rm cm}^{-2}$) for abundances $\sim$0.2 solar 
(SMC abundances) has been derived from simulations as

\begin{equation}
  H\!R1 = 1.00 - 0.82 \big(\frac{N_{\rm H}}{0.25}\big)^{-1.35} (N_{\rm H} \ge 0.3)
\end{equation}

\begin{equation}
  H\!R2 = 0.145 + 0.65 \big(\frac{N_{\rm H}}{15}\big)^{0.68} (N_{\rm H} \ge 0.3)
\end{equation}

\noindent
and for radio quiet AGN spectra (powerlaw photon index $-2.6$)

\begin{equation}
  H\!R1 = 1.00 - 1.6 \big(\frac{N_{\rm H}}{0.25}\big)^{-1.52} (N_{\rm H} \ge 0.3)
\end{equation}
\begin{equation}
  H\!R2 = 0.01 + 0.84 \big(\frac{N_{\rm H}}{15}\big)^{0.82} (N_{\rm H} \ge 0.3)
\end{equation}

For X-ray binaries with powerlaw photon index $-0.8$ spectra the dependence
of the X-ray colors $H\!R1$ and $H\!R2$ on the hydrogen column density is

\begin{equation}
  H\!R1 = 1.00 - 0.20 \big(\frac{N_{\rm H}}{0.25}\big)^{-0.94} (N_{\rm H} \ge 0.3)
\end{equation}

\begin{equation}
  H\!R2 = 0.415 + 0.48 \big(\frac{N_{\rm H}}{15}\big)^{0.72} (N_{\rm H} \ge 0.3)
\end{equation}

Equations~3 to 8 can be solved for the hydrogen column density $N_{\rm H}$.
In combining the equations for $H\!R1$ and $H\!R2$ one can derive analytical
solutions of X-ray binary and AGN tracks in the $H\!R1$ -- $H\!R2$ plane. 

For radio loud AGN (powerlaw photon index $-2.0$) one derives the track

\begin{equation}
  H\!R2 = 0.145 + 0.0402 \big(\frac{1.0 - H\!R1}{0.82}\big)^{-0.504} (N_{\rm H} \ge 0.3)
\end{equation}

and for radio quiet AGN (powerlaw photon index $-2.6$) the track

\begin{equation}
  H\!R2 = 0.01 + 0.0292 \big(\frac{1.0 - H\!R1}{1.6}\big)^{-0.539} (N_{\rm H} \ge 0.3)
\end{equation}

For X-ray binaries (powerlaw photon index $-0.8$) we derive the track

\begin{equation}
  H\!R2 = 0.415 + 0.0252 \big(\frac{1.0 - H\!R1}{0.16}\big)^{-0.766} (N_{\rm H} \ge 0.3)
\end{equation}

These tracks allow a classification of a {\sl ROSAT} source as an AGN
(or an X-ray binary) without the requirement that the intervening hydrogen 
column density is known. It even is possible to constrain in the 
$H\!R1$ -- $H\!R2$ plane the intervening hydrogen column density for the source. 
We note that X-ray binaries are not necessarily seen through the total gas
column of the galaxy disk. With a galactic foreground column of 
$4\times10^{20}\ {\rm cm^{-2}}$ towards NGC~3109 we derive with Equation~11 
for X-ray binaries a lower bound on $H\!R2$ of $H\!R2\ge0.45$.

The tracks given above have been derived for a metallicity $\sim$0.2 solar
which is consistent with the metallicity derived for H{\sc ii} regions in
NGC~3109 (cf. Minniti et al. 1999). Higher metallicity tracks deviate 
somewhat from these tracks especially if one approaches the high column 
density regime $H\!R1\sim1.0$.

\section{The H{\sc i} extent of NGC~3109} 

Huchtmeier et al. (1980, hereafter HSM80) measured the H{\sc i} distribution 
across NGC~3109 with the Effelsberg radio telescope and a beam size of 
9\arcmin. We apply the procedure used for the X-ray detections in the field
of the SMC to the X-ray detections in the field of NGC~3109. First we make 
use of the H{\sc i} map of HSM80 which has an extent of 60\arcmin\ $\times$ 
30\arcmin. We convert the H{\sc i} intensity into hydrogen column densities 
with the equation given in Dickey \& Lockman (1990), cf. Kahabka (1999) and 
we derive a peak of $N_{\rm H}\sim1.3\times10^{21}\ {\rm cm}^{-2}$. HSM80 
have found that NGC~3109 has a large extent in H{\sc i}, with an extension 
(distortion) of the H{\sc i} in the SW. With our X-ray catalog we cover the 
whole extent of the H{\sc i} of NGC~3109. If we restrict the analysis to the 
inner 20\arcmin\ of this field and to sources with well constrained hardness 
ratios $\delta H\!R2\le$0.2 then we can classify 7 sources as AGN.

\begin{figure}[htbp]
  \centering{ 
  \vbox{\psfig{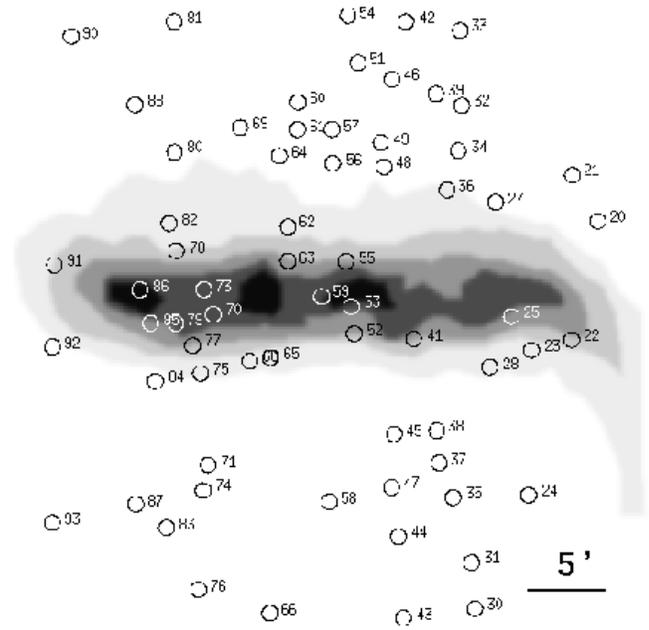}}\par
            }
  \caption[]{Positions of {\sl ROSAT PSPC} X-ray sources (the labeled 
            circles) overlaid on an H{\sc i} image of NGC~3109 obtained 
            with the VLA (JC90) as shown in gray scale. The maximum 
            hydrogen column density is $\rm 2.3\times10^{21}\ cm^{-2}$ and 
            the lowest hydrogen column density is $\rm 10^{19}\ cm^{-2}$. 
            \label{ps:hijc}}
\end{figure}

A much higher resolution H{\sc i} map of NGC~3109 with a beam size of 
40\arcsec\ has been derived by Jobin \& Carignan (1990, hereafter JC90) with 
the VLA. We now make use of this high-resolution H{\sc i} image which we take 
from plate~67 of JC90. The H{\sc i} distribution of NGC~3109 has an extent of 
40\arcmin\ $\times$ 12\arcmin. The peak hydrogen column density is 
$\rm \sim2.3\times10^{21}\ {\rm cm}^{-2}$, and the lowest column density 
is $\rm 10^{19}\ {\rm cm}^{-2}$. 

\begin{figure}[htbp]
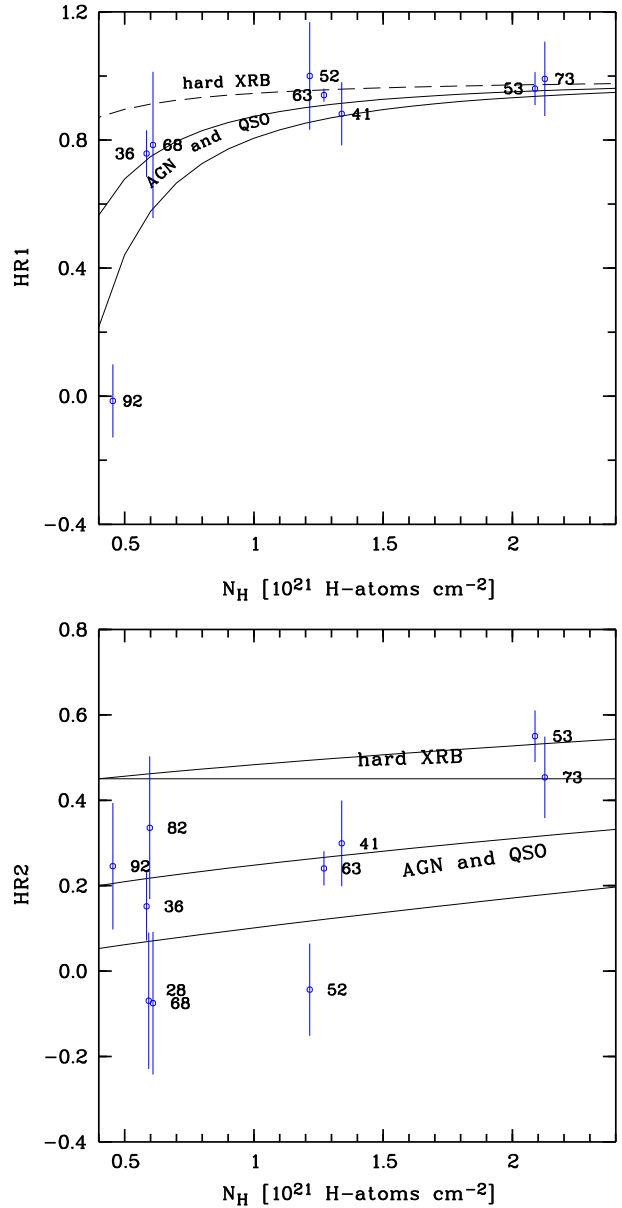

  \centering{ 
  \vbox{\psfig{figure=kahabka.f4a,width=8.8cm,%
  bbllx=3.7cm,bblly=1.5cm,bburx=16.5cm,bbury=13.5cm,clip=}}\par
  \vbox{\psfig{figure=kahabka.f4b,width=8.8cm,%
  bbllx=3.7cm,bblly=1.5cm,bburx=16.5cm,bbury=13.5cm,clip=}}\par
            }
  \caption[]{Hydrogen absorbing column density -- hardness ratio $H\!R1$ 
             (upper panel) and $H\!R2$ (lower panel) plane for 10 X-ray 
             sources within 20\arcmin\ of NGC~3109 with well determined 
             $H\!R1$ ($\delta H\!R1\le$0.24) and $H\!R2$ 
             ($\delta H\!R2\le$0.24). The hydrogen column density is the 
             NGC~3109 intrinsic and galactic foreground value. For NGC~3109 
             the H{\sc i} model of JC90 is used. Also given are the 
             theoretical curves for X-ray binaries (powerlaw slope $-0.8$) 
             and AGN (powerlaw slope $-2.0$ to slope $-2.6$) assimung a 
             metallicity $\sim$0.2 solar.
             \label{ps:hihr2jc}}
\end{figure}

\begin{figure}[htbp]
  \centering{ 
  \vbox{\psfig{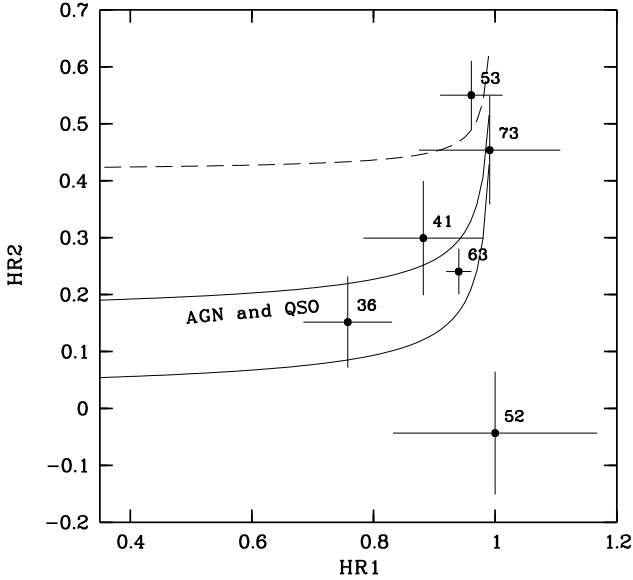}}\par
            }
  \caption[]{$H\!R1$ -- $H\!R2$ plane for 6 {\sl ROSAT} sources projected 
             onto the H{\sc i} extent of NGC~3109 and for which accurate 
             values $\delta H\!R1\le0.2$ and $\delta H\!R2\le0.2$ have been
             determined. Also shown are the two AGN tracks for powerlaw 
             photon index $-2.0$ (upper track) and $-2.6$ (lower track)
             and the X-ray binary track (dashed line).\label{ps:hr1hr2}}
\end{figure}

In Fig.~{\ref{ps:hijc}} we show the positions of the cataloged 
{\sl ROSAT} {\sl PSPC} sources overlaid on the gray scale H{\sc i} 
image of NGC~3109 and taken from JC90. We find 26 {\sl ROSAT} {\sl PSPC} 
sources within the H{\sc i} contours of JC90. If we restrict the 
analysis to the inner 20\arcmin\ of this field and to sources with 
well constrained hardness ratios $\delta H\!R2\le$0.2, then we can 
classify 3 sources (number 36, 41 and 63) as AGN and 2 sources (number 
53 and 73) as X-ray binaries. Source 53 may be seen through higher gas 
columns while source 73 is seen through lower gas columns (see 
Fig.~\ref{ps:hihr2jc} and Equation~11). One source (number 82) can 
be either class. The two sources 52 and 92 cannot be classified as AGN 
or X-ray binaries. Source 92 may be a foreground object (cf. 
Fig.~\ref{ps:hihr2jc}, Fig.~\ref{ps:hr1hr2} and Table~\ref{tab:classjc}). 
Source 52 has similar hardness ratios as the LMC SNR 0548-70.4 (Haberl 
\& Pietsch 1999). It could be a (young) SNR in NGC~3109.

\begin{table}[htbp]
  \caption[]{The 9 classified sources in the central 20\arcmin\ of NGC~3109 
             using the H{\sc i} data of JC90. A = background AGN, B = X-ray 
             binary source, F foreground star, AB = class A or class B.}
  \begin{flushleft}
  \begin{tabular}{ll}
  \hline
  \noalign{\smallskip}
   Source No. & Class   \\ 
  \noalign{\smallskip}
  \hline
  \noalign{\smallskip}
   36, 41, 63                 & A   \\
  \noalign{\smallskip}
   53, 73                     & B   \\
  \noalign{\smallskip}
   82                         & AB  \\
  \noalign{\smallskip}
   28, 68, 92                 & F   \\
  \noalign{\smallskip}
  \hline
  \end{tabular}
  \end{flushleft}
  \label{tab:classjc}
\end{table}

NGC~3109 has a mass smaller than, or comparable to, the LMC and 2 X-ray 
binaries with luminosities above a few times $\sim 10^{36}\ {\rm erg\ s}^{-1}$
would be in agreement with extrapolations from {\sl ROSAT} findings for the 
LMC (cf. Haberl \& Pietsch 1999). 

There are 14 {\sl ROSAT PSPC} sources which are projected onto NGC~3109 
intrinsic hydrogen columns of $N_{\rm H}\ge 10^{21}\ {\rm cm}^{-2}$ (cf. 
Fig.~\ref{ps:hijc}). The largest column is derived for source 86 
($N_{\rm H} = 2.3\ 10^{21}\ {\rm cm}^{-2}$). This source could be associated 
with a spiral arm or an H{\sc ii} region of NGC~3109. Two further sources,
53 and 59, are close to another region of large column density and source 
59 is also close to the optical center of NGC~3109.

We further test the correctness of the classification by constructing the 
distribution of the number of detected X-ray sources $N$ with fluxes in 
excess of a given flux $S$, the $\log N\ -\ \log S$ (cf. Hasinger et al. 
1993).  We restrict the analysis to sources within the central 20\arcmin\ 
of NGC~3109 which have been classified as AGN or X-ray binaries. We correct 
the X-ray fluxes for the intervening hydrogen columns by using for the 
galactic contribution a value of $\rm 4.3\ 10^{20}\ cm^{-2}$ (Dickey \& 
Lockman 1990) and for the contribution due to NGC~3109 the H{\sc i} model 
of JC90. We find that the observed $\log N\ -\ \log S$ is in excess of
the $\log N\ -\ \log S$ of the soft extragalactic X-ray background 
(Hasinger et al. 1993). This can be accounted for if an additional less steep 
component (e.g. due to X-ray binaries) is added. If we only construct 
the $\log N\ -\ \log S$ of the candidate AGN, then the $\log N\ -\ \log S$ 
of the soft extragalactic X-ray background is reproduced. This would mean 
that no significant additional
hydrogen is needed to explain the observed sources as background AGN and X-ray
binaries although only a few candidate AGN are within the H{\sc i} extent of 
NGC~3109. The two brightest candidate AGN (source 36 and 63) could give an 
excess in the $\log N\ -\ \log S$. We cannot exclude that they are at least 
in part related to NGC~3109.

\section{Optical counterparts of the X-ray sources}

We generated finding charts in the B-band {\sl COSMOS} blue plates and
in the I-band ({\sl NTT EMMI}, cf. Fig.~\ref{ps:charts}) to search for 
optical counterparts of the {\sl ROSAT} sources. The B-band plates were 
used to classify the source using the flux ratio $f_x/f_{\rm opt}$ while 
the I-band image was used to determine the source morphologies from an 
analysis of their point spread function (see below).

\subsection{Matches in the B-band}

We produced finding charts of the {\sl ROSAT} sources given in 
Table~6 using the {\sl COSMOS} blue plates. We 
especially investigated the classified {\sl ROSAT} sources given in 
Table~\ref{tab:classjc}. 

We searched for the optical counterparts in the 90\% confidence circle
of the {\sl ROSAT} source. We list the parameters of the matches in 
Table~\ref{tab:matches}. For the two sources 28 and 92 we find bright 
(B=11.7~mag and B=12.1~mag) optical counterparts in the 90\% confidence 
circle of these {\sl ROSAT} sources. This indicates a galactic foreground 
nature of these sources. For the sources 36, 41, 52, 68 and 73 we find weak 
(B=20.4~mag to B=22.3~mag) unresolved optical counterparts. For source 73 
there is also a galaxy (B=22.7~mag) in the error circle. For source 53 we 
cannot find an optical counterpart as it is in the crowded field of NGC~3109 
and for 63 no optical counterpart exists within the {\sl ROSAT} 90\%
confidence circle.

\begin{table}[htbp]
  \caption[]{Optical ({\sl COSMOS} blue plates) matches for {\sl ROSAT}
             sources from Table~\ref{tab:classjc}. Given is the B-mag,
             $\log (f_{\rm x}/f_{\rm opt})$ and some rough source 
             classification (stellar, galaxy, unresolved or faint).}
  \begin{flushleft}
  \begin{tabular}{rccc}
  \hline
  \noalign{\smallskip}
   Source  & B-mag & $\log (f_{\rm x}/f_{\rm opt})$ & Comment         \\ 
   No.     &       &                               &                 \\
  \noalign{\smallskip}
  \hline
  \noalign{\smallskip}
   28       & 11.7  & --3.8                        & foreground star \\
   36       & 20.4  &  0.33                        & unresolved, AGN \\
   41       & 21.1  &  0.35                        & unresolved, AGN \\
   52       & 20.4  &  0.01                        & unresolved      \\
   63       &\multicolumn{3}{c}{no counterpart}                      \\
   68       & 22.3  & 0.47                         & unresolved      \\
            & 24.4  & 1.3                          & faint           \\
   73       & 21.7  & 0.56                         & unresolved      \\
            & 22.7  & 0.96                         & galaxy          \\
            & 25.9  & 2.2                          & faint           \\
   92       & 21.2  & 0.48                         & unresolved      \\
            & 12.1  & --3.2            & foreground star$^{\rm (a)}$ \\
  \noalign{\smallskip}
  \hline
  \end{tabular}
  \end{flushleft}
  Note (a): 17\arcsec\ distant from the {\sl ROSAT} source
  \label{tab:matches}
\end{table}

In addition we calculated the flux ratio $f_{\rm x} / f_{\rm opt}$ with
the equation given in Haberl \& Pietsch (1999). We make use of the 
B-magnitude given in Table~\ref{tab:matches} and the {\sl PSPC} countrate
given in Table~6. We use the abbreviation 
$k_{\rm PSPC} = \log (10^{-11}\ {\sl PSPC}\ {\rm countrate})$
and obtain the flux ratio from the equation:

\begin{equation}
  \log (f_{\rm x}/f_{\rm opt}) = k_{\rm PSPC} + 0.4\ m_{\rm B} + 5.37
\end{equation}

Using the scheme given in Table~3 of Haberl \& Pietsch (1999) we classify
the sources 28 and 92 as candidate foreground stars and the sources 36 and 
41 as candidate AGN (cf. Table~\ref{tab:matches}).

\subsection{Matches in the I-band}

\begin{figure*}[htbp]
  \centering{ 
  \vbox{\psfig{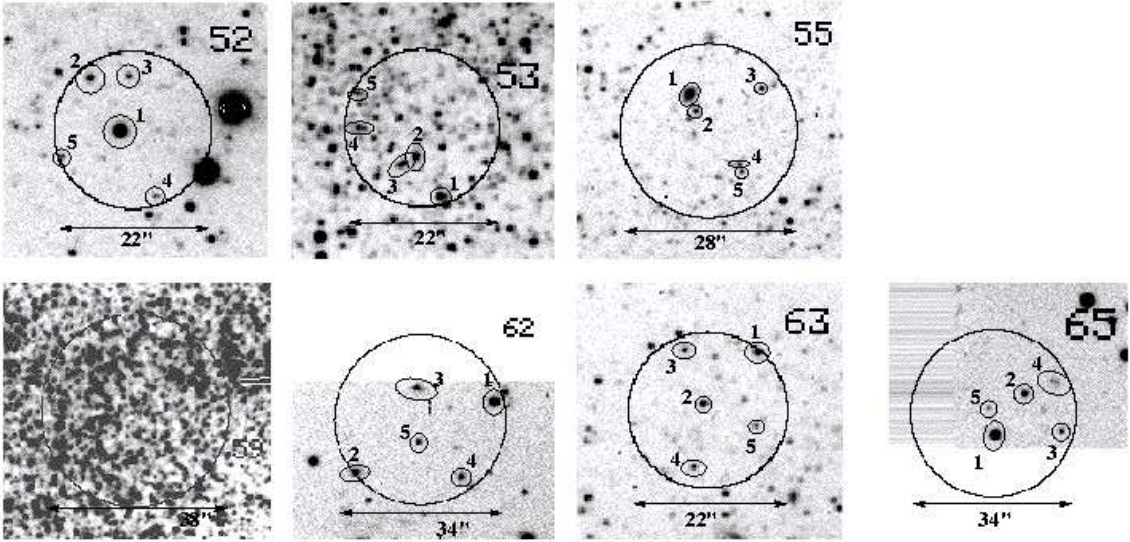}}\par
            }
  \caption[]{Finding charts for 7 {\sl ROSAT PSPC} sources. The
    nomenclature of the charts is consistent with the source number
    in Table~6.}
  \label{ps:charts}
\end{figure*}

For the first run of optical matching of X-ray sources in the central
part of NGC~3109 we obtained three optical I-band images (60.E-0818)
from the ESO archive.\footnote{Based on observations made with ESO 
Telescopes at the La Silla or Paranal Observatories under programme 
ID 60.E-0818(A).} All images were taken on 1998 February 2 with
the red arm of the ESO Multi-Mode Instrument ({\sl EMMI}) at the New
Technology Telescope ({\sl NTT}) with a total exposure time of 2700 s.  The
images were taken with the Tektronix 2048$\times$2048 pix$^2$ chip
with a scale of 0.268\arcsec /pix. The resulting field of view covered
by the stacked images is 8.8\arcmin\ $\times$ 8.8\arcmin\ and the final
image is centered on the galaxy nucleus. The seeing was measured to be
$\sim0.75$\arcsec\ throughout all three exposures.

The data are used to identify optical couterparts within the region of
highest H{\sc i} column density, which resides at the center of NGC~3109
(Jobin \& Carignan 1990) and is best covered by the field of view of
this particular dataset, and to support or reject the classification
provided by the hardness-ratio estimate (see Sect.~5). Further
photometric studies in optical passbands of {\sl ROSAT} sources for most
objects in the catalog in Table~6 with accurate positions
will be given in a subsequent paper.

In order to obtain bona-fide optical counterparts we transformed the
pixel coordinates of the I-band to new equatorial coordinates.
We used positions of 43 stars out of the USNO
Astrometric Catalog (Monet et al. 1998) which was obtained from
Centre de Donn\'ees de Astronomiques de Strasbourg (CDS). Using the
task {\it ccmap} within the IRAF\footnote{IRAF is distributed by the
  National Optical Astronomy Observatories, which are operated by the
  Association of Universities for Research in Astronomy, Inc., under
  cooperative agreement with the National Science Foundation.}
environment (Tody et al. 1993) we find the mean rms uncertainty for
all given optical coordinates being $\Delta$RA$\leq 0.62$\arcsec\ and
$\Delta$Dec$\leq 0.48$\arcsec. Applying the plate solution to
coordinates provided by the {\sl ROSAT PSPC} catalog (see Table~6) 
we find 7 optical matches within the field of
view. The finding charts for all these regions are given in 
Fig.~\ref{ps:charts} with the appropriate scaling of the 90\%-confidence
circle as given in Table~6. The five brightest objects 
are encircled and labeled according to their luminosity (label~1 marks 
the brightest object).

\begin{figure}[htbp]
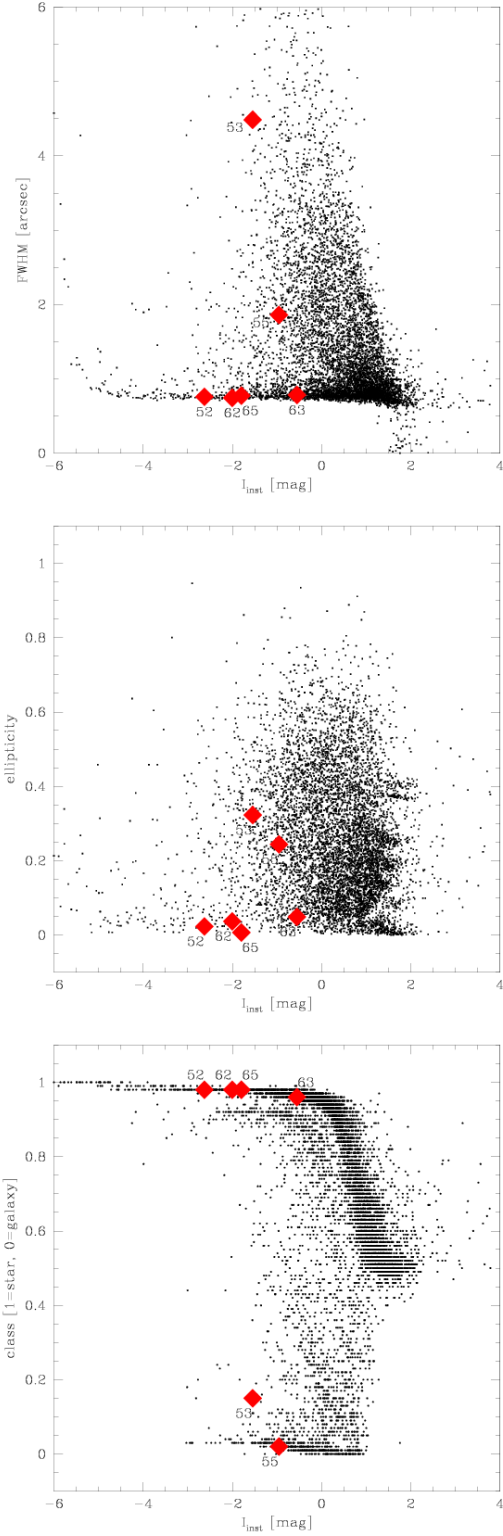

  \centering{ 
  \vbox{\psfig{figure=kahabka.f7a,width=6.8cm,%
  bbllx=0.0cm,bblly=0.0cm,bburx=19.5cm,bbury=19.7cm,clip=}}\par
  \vbox{\psfig{figure=kahabka.f7b,width=6.8cm,%
  bbllx=0.0cm,bblly=0.0cm,bburx=19.5cm,bbury=19.7cm,clip=}}\par
  \vbox{\psfig{figure=kahabka.f7c,width=6.8cm,%
  bbllx=0.0cm,bblly=0.0cm,bburx=19.5cm,bbury=19.7cm,clip=}}\par
          }
  \caption[]{Optical shape parameters for all optical sources in the
    {\sl EMMI} I-band image. We added (with number) the PSF-shape 
    parameters of each brightest object within the finding circles in 
    Fig.~\ref{ps:charts}. The upper panel shows the FWHM versus 
    instrumental I-magnitude. The middle panel gives the ellipticity 
    distribution as a function of I-band magnitude while the lower panel 
    is the star-galaxy-classification (0 being an extended and 1 being 
    a point source) versus I.\label{ps:shapepars}}
\end{figure}

Subsequently, we perform an analysis of the shape parameters
of the optical point spread function (PSF) for the 5 brightest objects
within each of the 7 object's finding circles. We used the
source-extraction software SExtractor v2.1.4 (Bertin \& Arnouts 1996)
for determination of ellipticity $\varepsilon$, full width at half
maximum (FWHM), and the star-galaxy-classification parameter CL (using
a neural-network algorithm which was extensively trained, see Bertin
\& Arnouts for details). We compare our findings to all remaining
optical objects most of which are assumed to be point-sources. The
code yields $\sim$7500 detections for which we plot the three
PSF-shape parameters ($\varepsilon$, FWHM, and CL) as a function of
instrumental I-band magnitude. Fig.~\ref{ps:shapepars} shows the
parameters of all detections together with the parameters for the
brightest source within the finding circle (see Table~4). Note that the 
brightest object is {\it not} necessarily the optical counterpart of 
the {\sl ROSAT} source.

\begin{table}[htbp]
  \caption[]{Shape parameters of the 5 brightest sources within 90\%
    confidence finding circle of 7 optically matched {\sl ROSAT} source
    coordinates. Col.~1 gives the number which agrees with the
    labels in Fig.~\ref{ps:charts}. Col.~2 and 3 are the instrumental
    magnitude and the corresponding error. The FWHM, ellipticity, and
    the star-galaxy-parameter (from 1=star to 0=galaxy) as
    calculated by SExtractor are given in col.~4, 5, and 6.} 
  \begin{flushleft}
  \begin{tabular}{crrrrr}
  \hline
  \noalign{\smallskip}
Label$^a$& I$_{\rm inst}\;^b$ & $\Delta$I$_{\rm inst}\;^b$ & FWHM$^c$ & ell$^d$ & class$^e$ \\\hline\hline
\noalign{\smallskip}
\noalign{\smallskip}
\multicolumn{6}{c}{source 52}\\\hline
1  & $-2.63$&    0.01  &     0.76 &     0.023 &      0.98 \\
2  &   0.01 &    0.03  &     0.90 &     0.076 &      0.96 \\
3  &   0.80 &    0.05  &     0.81 &     0.024 &      0.84 \\
4  &   0.83 &    0.06  &     0.93 &     0.089 &      0.77 \\
5  &   0.93 &    0.06  &     0.79 &     0.065 &      0.66 \\
\noalign{\smallskip}
\noalign{\smallskip}
\multicolumn{6}{c}{source 53}\\\hline
1  & $-1.54$&    0.01  &     4.52 &     0.323 &      0.15 \\
2  & $-0.86$&    0.01  &     0.87 &     0.337 &      0.96 \\
3  & $-0.68$&    0.01  &     1.36 &     0.229 &      0.97 \\
4  & $-0.59$&    0.02  &     7.17 &     0.373 &      0.02 \\
5  & $-0.59$&    0.02  &     5.69 &     0.622 &      0.01 \\
\noalign{\smallskip}
\noalign{\smallskip}
\multicolumn{6}{c}{source 55}\\\hline
1  &$ -0.95$&    0.01  &     1.87 &     0.244 &      0.02 \\
2  &   0.05 &    0.03  &     2.68 &     0.448 &      0.17 \\
3  &   0.39 &    0.04  &     0.75 &     0.065 &      0.88 \\
4  &   0.60 &    0.05  &     1.42 &     0.301 &      0.81 \\
5  &   0.79 &    0.05  &     0.87 &     0.231 &      0.73 \\
\noalign{\smallskip}
\noalign{\smallskip}
\multicolumn{6}{c}{source 59}\\\hline
\multicolumn{6}{c}{too crowded, no reliable parameters can be given}\\
\noalign{\smallskip}
\noalign{\smallskip}
\multicolumn{6}{c}{source 62}\\\hline
1  &$ -2.00$&    0.01  &     0.75 &     0.037 &      0.98 \\
2  &$ -0.23$&    0.02  &     0.95 &     0.066 &      0.81 \\
3  &   0.01 &    0.03  &     1.41 &     0.288 &      0.81 \\
4  &   0.20 &    0.03  &     1.16 &     0.102 &      0.78 \\
5  &   0.65 &    0.05  &     1.00 &     0.015 &      0.84 \\
\noalign{\smallskip}
\noalign{\smallskip}
\multicolumn{6}{c}{source 63}\\\hline
1  &$ -0.55$&    0.02  &     0.79 &     0.049 &      0.96 \\
2  &$ -0.52$&    0.02  &     0.73 &     0.059 &      0.96 \\
3  &   0.21 &    0.03  &     0.81 &     0.088 &      0.89 \\
4  &   0.47 &    0.04  &     0.72 &     0.076 &      0.86 \\
5  &   0.97 &    0.06  &     0.77 &     0.075 &      0.67 \\
\noalign{\smallskip}
\noalign{\smallskip}
\multicolumn{6}{c}{source 65}\\\hline
1  & $-1.80$&    0.01  &     0.78 &     0.007 &      0.98 \\
2  & $-0.52$&    0.02  &     0.80 &     0.033 &      0.96 \\
3  &   0.80 &    0.05  &     0.82 &     0.011 &      0.62 \\
4  &   1.03 &    0.07  &     1.85 &     0.414 &      0.67 \\
5  &   1.77 &    0.13  &     0.72 &     0.245 &      0.49 \\
  \noalign{\smallskip}
  \hline
  \noalign{\smallskip}
  \end{tabular}
  \end{flushleft}
   (a) Numbering in agreement with labels in Fig.~\ref{ps:charts}
   (b) Instrumental I magnitude and the photometric error in mag.
   (c) Full width at half maximum in arc seconds.
   (d) Ellipticity, i.e. 1--a/b.
   (e) star-galaxy-classification parameter of SExtractor; objects
   with CL$=1$ are point sources while CL$=0$ classifies objects as extended.
\label{tab:shapepars}
\end{table}

The brightest sources 52.1, 62.1 and 65.1 (the first number being the source
number and the decimal the source label) in Fig.~\ref{ps:charts} are at
least one magnitude brighter than the next fainter object within the finding 
circle (see Table~\ref{tab:shapepars}). Object 63.1 has about the same 
magnitude as object 63.2. Therefore 
no clear preference can be given just from considering the I-band flux. 
The PSF-shape parameters of these 4 objects agree with those found for stars 
(see Fig.~\ref{ps:shapepars}). The CL classification of SExtractor puts 
all 4 sources in the point-source regime. Yet, the optical findings are
prone to mis-classification since not every brightest object is most
central within the finding circle. 

Although they are not the brightest, the sources 62.5 and 63.2 are the most 
likely candidate optical counterparts (infering from their most central 
position only). These were classified as slightly extended and slightly 
elliptical, respectively.

As can be seen from Fig.~\ref{ps:charts} there are 2 finding charts
(i.e.  charts 62 and 65) from the edge of the CCD image. Thus, we cannot
exclude brighter objects within the area not covered by the exposure.
Chart 59 is too crowded to give a reliable identification only on
the basis of any photometric PSF-shape parameter. However, we note that the 
center of the {\sl ROSAT} error circle coincides with the galaxy's center. 
Thus, it is very likely that the {\sl ROSAT} source is connected with a 
X-ray source which is located near the center of the galaxy.

\section{Candidate AGN}

In Sect.5 we have used the hardness ratios $H\!R1$ and $H\!R2$ to 
classify 3 {\sl ROSAT} sources (with numbers 36, 41 and 63) and projected 
onto the H{\sc i} extent of NGC~3109 as candidate AGN. 

We have searched in the optical B-band {\sl COSMOS} finding charts of these 
sources for optical counterparts. In addition we have generated deep 
I-band finding charts for source 63 and other {\sl ROSAT} sources and we 
have looked for faint unresolved and extended objects within the positional 
error circles of these {\sl ROSAT} sources (cf. Fig.~\ref{ps:charts}). 
We have found several optical candidates in the positional error circle of 
each {\sl ROSAT} source. Therefore optical spectroscopy of all candidates is 
required to firmly identify the optical counterpart of these {\sl ROSAT} 
sources. 

\begin{figure}[htbp]
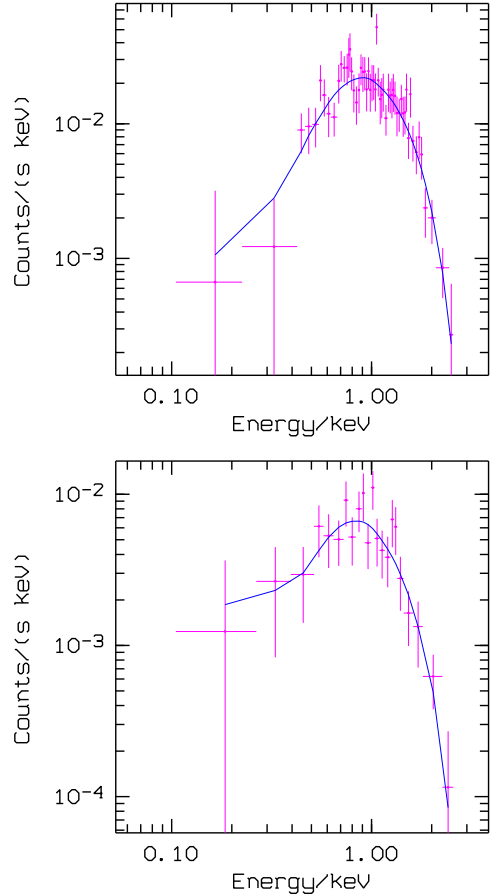

  \centering{ 
  \vbox{\psfig{figure=kahabka.f8a,width=6.8cm,%
  bbllx=1.5cm,bblly=7.0cm,bburx=13.5cm,bbury=17.7cm,clip=}}\par
  \vbox{\psfig{figure=kahabka.f8b,width=6.8cm,%
  bbllx=1.5cm,bblly=7.0cm,bburx=13.5cm,bbury=17.7cm,clip=}}\par
            }
  \caption[]{{\sl ROSAT PSPC} X-ray spectra and best-fit spectral models 
             (assuming galactic and NGC~3109 intrinsic absorption) of the 
             AGN RX~J1003.2-2607 (upper panel) and RX~J1002.5-2602 (lower 
             panel).\label{ps:xrayfit}}
\end{figure}

If the background nature of candidate AGN ``close to'' NGC~3109 is
established then these AGN can be used to probe the gas column density of 
NGC~3109 in the direction of these AGN (taking the additional galactic 
gas column into account). 
For two sufficiently X-ray bright candidate AGN RX~J1003.2-2607 (source 63) 
and RX~J1002.5-2602 (source 36) we perform X-ray spectral fitting with the 
{\sl EXSAS} spectral analysis task (Zimmermann et al. 1994). We assume a
galactic foreground column density of $4.3\ 10^{20}\ {\rm cm}^{-2}$ and 
determine the additional hydrogen column density (due to NGC~3109) assuming 
reduced metallicities ($\sim$0.2 solar). For both sources we find that the 
hydrogen column deduced from the X-ray spectral fit is consistent with the 
hydrogen column deduced from the 21-cm survey of HSM80 and JC90 (cf. 
Table~\ref{tab:specpar} and Fig.~\ref{ps:xrayfit}).

\begin{table}[htbp]
  \caption[]{NGC~3109 intrinsic hydrogen column densities in the direction
             of the candidate AGN RX~J1003.2-2607 and RX~J1002.5-2602 as 
             deduced from X-ray spectral fitting. Also given is the hydrogen
             column density derived from the 21-cm data of HSM80 and CJ90. 
             68\% confidence errors are given. A galactic hydrogen column 
             density of $\rm 4.3\times10^{20}\ cm^{-2}$ and a metallicity 
             $\sim$0.2 solar is assumed in the X-ray spectral fit. 
             Also given is the spectral photon index $\rm \alpha$ and the 
             unabsorbed X-ray flux (0.1 -- 2.0 keV).}
  \begin{flushleft}
  \begin{tabular}{cccccc}
  \hline
  \noalign{\smallskip}
  $\rm \alpha$&flux ($\rm 10^{-13}$ &\multicolumn{4}{c}{hydrogen column}    \\
  &$\rm erg\ cm^{-2}\ s^{-1}$)&\multicolumn{4}{c}{($\rm 10^{20}\ cm^{-2}$)}\\
  &    &\multicolumn{2}{c}{X-ray}       &\multicolumn{2}{c}{21-cm}          \\
  &    & (a)     & (b)  & (c) & (d)                \\ 
  \noalign{\smallskip}
  \hline
  \noalign{\smallskip}
  \multicolumn{6}{c}{RX~J1003.2-2607  (source 63)} \\
  \noalign{\smallskip}
  -2.2$\pm^{0.2}_{0.6}$ & 7.9$\pm^{11}_{3.2}$ & 11$\pm^{15}_{5}$  &11$\pm^{7}_{5}$ & 10 & 8$\pm$2 \\
  \noalign{\smallskip}
  \hline
  \noalign{\smallskip}
  \multicolumn{6}{c}{RX~J1002.5-2602  (source 36)} \\
  \noalign{\smallskip}
  -2.4$\pm^{0.6}_{0.9}$ & 3.1$\pm^{17}_{1.7}$ & 4.1$\pm^{21}_{4}$ &4.1$\pm^{6}_{4}$& 4 & 1 \\
  \noalign{\smallskip}
  \hline
  \noalign{\smallskip}
  \end{tabular}
  \end{flushleft}
   (a) No constraint on $\rm \alpha$;
   (b) constraint on powerlaw photon index: $\rm -1.75\le \alpha \le -2.25$;
   (c) H{\sc i} model of HSM80;
   (d) H{\sc i} model of JC90.
   \label{tab:specpar}
\end{table}

This result is also consistent with the location of these two sources in 
the $H\!R1$ -- $H\!R2$ plane (cf. Fig.~\ref{ps:hr1hr2}).
We do not include in the spectral analysis source 53 which has been 
classified as a candidate X-ray binary in Sect.5.  

\section{Molecular content of NGC~3109}

Assuming that both AGN have canonical powerlaw photon indices $\rm \alpha$ 
= $-$(1.75 to 2.25), cf. Laor et al. (1997), we further tighten the 
constraints on the hydrogen column density. For RX~J1003.2-2607 we find from 
the X-ray spectral fit $N_{\rm H} = 11\pm^{7}_{5}\times10^{20}\ 
{\rm cm}^{-2}$ and from the 21-map of JC90 $N_{\rm H} = 8\pm2\times10^{20}\
{\rm cm}^{-2}$. For RX~J1002.5-2602 we constrain the hydrogen column density 
from the X-ray spectral fit to $N_{\rm H} = 4\pm^{6}_{4}\times10^{20}\ 
{\rm cm}^{-2}$ and from the 21-map of JC90 to $N_{\rm H}\sim1\times10^{20}\ 
{\rm cm}^{-2}$. Both values are in agreement within the uncertainties.

The column densities of absorbing material from the X-ray data are slightly
larger than those from the 21-cm line, but the uncertainty of the X-ray 
spectral fit leaves room for some additional hydrogen (e.g. molecular 
hydrogen) in the line of sight of the AGN.
We assume that the photoionisation cross section is $\sim2.8$ times larger 
for molecular hydrogen than for atomic hydrogen (cf. Cruddace et al. 1974; 
Yan et al. 1998). Then we determine the column density due to molecular 
hydrogen $N_{\rm H_{2}}$ from the total hydrogen column density 
$N_{\rm H}^{\rm tot}$ derived from the X-ray spectral fit and the atomic 
hydrogen column density $N_{\rm H{\sc i}}$ derived from the 21-cm 
observations as

\begin{equation}
N_{\rm H_{2}} = \frac{1}{2.8} \big(N_{\rm H}^{\rm tot} - N_{\rm H{\sc i}}\big) 
\end{equation}

\noindent
We so constrain the gas column due to molecular hydrogen in NGC~3109 from 
the X-ray spectral fit of RX~J1003.2-2607 to 
$\approxlt4\times10^{20}\ {\rm cm^{-2}}$.

Towards RX~J1003.2-2607 the amount of $\rm H_{2}$ is $\sim$$10^{20}\ 
{\rm cm}^{-2}$. Along this line of sight the molecular mass fraction, 
$f_{\rm m} = (N_{\rm H}^{\rm tot} - N_{\rm H{\sc i}}) / (N_{\rm H}^{\rm tot} 
+ 0.4\ N_{\rm H{\sc i}}) = 0.21\pm0.37$. This means $\approxlt60$\% of the 
mass of the total gas is in molecular form. This result can be compared with 
CO observations of NGC~3109, from which the mass of molecular hydrogen has 
been determined to $\approxgt$$4\ 10^7\ M_{\odot}$ (Rowan-Robinson, Philips \& 
White 1980). With a H{\sc i} mass of $5\ 10^8\ M_{\odot}$ a molecular mass 
fraction of $\approxgt$10\% is obtained.

Extraplanar absorbing clouds have been found in a high-resolution survey of 
a sample of 12 edge-on galaxies by Howk \& Savage (1999) at distances of 
$z~=~0.5$ to $1.5$~kpc from the galaxy plane. For an inclination of 80\D\, a 
radius of the galaxy disk of 12~kpc and a distance to the galaxy of 1.4~Mpc 
similar clouds would be projected in NGC~3109 1\arcmin\ to 4\arcmin\ 
from the galaxy plane. From these clouds dust absorption has been observed
but the dust should coexist with gas in the molecular and atomic phase.   
Similar clouds may exist in NGC~3109 and the candidate AGN RX~J1003.2-2607 
would be seen through gas which is at a height $z\le$1~kpc above the galaxy 
plane while the candidate AGN RX~J1002.5-2602 is seen through gas with 
$z\le$2.4~kpc.

\section{Summary and conclusions}

From {\sl ROSAT} {\sl PSPC} observations of the dwarf galaxy NGC~3109 we
derive 10 X-ray sources which are contained within the H{\sc i} extent of 
this galaxy and which have accurate hardness ratios. We classify 2 of these
sources as foreground stars, 2 as candidate X-ray binaries and 3 as candidate 
background AGN. From X-ray spectral fitting we derive for 2 of the AGN total 
hydrogen column densities which we compare with the H{\sc i} column densities
inferred from 21-cm line measurements. We estimate that the molecular mass
fraction of the gas is not larger than $\sim$60\%.
Upcoming spectroscopy of candidate optical counterparts will help to constrain
the nature of these objects and help to understand the hydrogen content of 
NGC~3109. Furthermore, spectroscopy of globular clusters around NGC~3109 will 
help to trace the enrichment history (i.e. the major star formation episodes) 
and the kinematics (i.e. total mass) of NGC~3109.

\acknowledgements
The \ros\ project is supported by the Max-Planck-Gesellschaft and the 
Bundesministerium f\"ur Forschung und Technologie (BMFT). This research made 
use of the {\sl COSMOS} digitized optical survey of the southern sky, 
operated by the Royal Observatory Edinburgh and the Naval Research Laboratory,
with support from NASA. This research has made use of the SIMBAD data base
operated at CDS, Strasbourg, France. We thank K.S. de Boer and U. Klein 
for critically reading the manuscript. We thank an anonymous referee for
useful comments.

\clearpage
\vfill \eject


\begin{thebibliography}{}
\bibitem{} Bertin E, Arnouts, S., 1996, A\&AS 117, 393
\bibitem{} Brinkmann W., Yuan W., \& Siebert J., 1997, A\&A 319, 413
\bibitem{} Carignan C., 1985, ApJ 299, 59
\bibitem{} Cash W., 1979, ApJ 228, 939  
\bibitem{} Cruddace R., Paresce F., Bowyer S., \& Lampton M., 1974, 
           ApJ 187, 497 
\bibitem{} Cruddace R.G., Hasinger G., \& Schmitt J.H.M.M., 1988, in: 
           Astronomy from Large Databases, eds. Murtagh F., Heck A., p. 177
\bibitem{} Demers S., Irwin M.J., \& Kunkel W.E., 1985, AJ 90, 1967
\bibitem{} Dickey J.M., \& Lockman F.J., 1990, ARA\&A 28, 215
\bibitem{} Haberl F., \& Pietsch W., 1999, A\&AS 139, 277
\bibitem{} Hasinger G., Burg R., Giacconi R., et al., 1993, A\&A 275, 1
\bibitem{} Howk J.C., \& Savage B.D., 1999, AJ 117, 2077
\bibitem{} Huchtmeier W.K., Seiradakis J.H., \& Materne J., 1980, A\&A 91,
           341 [HSM80]
\bibitem{} Jobin M., \& Carignan C., 1990, AJ 100, 648 [JC90]
\bibitem{} Kahabka P., 1999, A\&A 344, 459 
\bibitem{} Kahabka P., Pietsch W., Filipovic M.D., \& Haberl F., 1999,
           A\&AS 136, 81 [KPFH99]
\bibitem{} K\"urster M., 1993, ROSAT Status Report No. 67 
\bibitem{} Laor A., Fiore F., Elvis M., et al., 1997, ApJ 477, 93
\bibitem{} Mateo M., 1998, ARA\&A 36, 1998
\bibitem{} Materne J., 1990, {\it ESO/ESA Workshop on Dwarf Galaxies},
           ed. M. Taringhi \& K. Kjar, ESO Garching, p.67
\bibitem{} Minniti D., Zulstra A.A., Alonso M.V., 1999, AJ 117, 881
\bibitem{} Monet D., Bird A., Canzian B., Dahn C., Guetter H., Harris
  H., Henden A., Levine S., Luginbuhl C., Monet A.K.B., Rhodes A.,
  Riepe B., Sell S., Stone R., Vrba F., Walker R., 1998, U.S. Naval
  Observatory Flagstaff Station (USNOFS) and Universities Space
  Research Association (USRA) stationed at USNOFS.
\bibitem{} Musella I., Piotto G., \& Capaccioli M., 1997, AJ~114, 976
\bibitem{} Pfeffermann E., Briel U.G., Hippmann H., et al., 1986, Proc. SPIE 
           733, 519
\bibitem{} Richer M.G., \& McCall M.L., 1995, ApJ~445, 642
\bibitem{} Rowan-Robinson M., Philips T.G., \& White G., 1980, A\&A 82, 381
\bibitem{} Sandage A., \& Tammann G.A., 1981, {\it A Revised Shapley-Ames
           Catalogue of Bright Galaxies}, Carnegie Institurion of Washington,
           Washington D.C.
\bibitem{} Stanimirovic S., Staveley-Smith L., Dickey J.M., et al., 1999
           MNRAS 302, 417 
\bibitem{} Tody, D. 1993, "IRAF in the Nineties" in Astronomical Data
  Analysis Software and Systems II, A.S.P. Conference Ser., Vol 52,
  eds. R.J. Hanisch, R.J.V. Brissenden, \& J. Barnes, 173. 
\bibitem{} Tr\"umper J., 1983, Adv. Space Res. 2, No. 4, 241
\bibitem{} van den Bergh S., 1999, ApJ 517, L97
\bibitem{} Vennes S., 1999, ApJ 525, 995
\bibitem{} Voges W., Aschenbach B., Boller T., et al., 1999, A\&A 349, 389 
\bibitem{} Whiting A.B., Irwin M.J., \& Hau G.K.T., 1997, AJ 114, 996
\bibitem{} Yan M., Sadbeghpour H.R., \& Dalgarno A., 1998, ApJ 496, 1044 
\bibitem{} Zimmermann H.U., Becker W., Belloni T., et al., 1994, 
           MPE report 257
\end{thebibliography}
\end{document}